\documentclass[global]{svjour}
\usepackage{latexsym}
\usepackage{color}
\usepackage{graphics}
\journalname{myjournal}
\begin{document}
\title{A $\rm{1.82~m^2}$  ring laser gyroscope for nano-rotational motion sensing}
\author{Jacopo~Belfi\inst{1}, Nicol\`o~Beverini\inst{1}, Filippo Bosi\inst{2},
Giorgio~Carelli\inst{1}, Angela Di Virgilio\inst{2}, Enrico~Maccioni\inst{1}, Antonello~Ortolan\inst{3} and Fabio~Stefani\inst{1}}
%
%
\institute{Department of Physics ``Enrico Fermi'', Universit\`a di Pisa,
and\\ CNISM unit\`a di Pisa, Italy \and INFN Sez. di Pisa, Pisa, Italy \and Laboratori Nazionali di Legnaro, INFN Legnaro (Padova), Italy}
\date{Received: date / Revised version: date}
\authorrunning{J. Belfi et al.}
\maketitle
\begin{abstract}
We present a fully active-controlled He-Ne ring laser gyroscope operating in square cavity 1.35~m in side. The apparatus is designed to provide a very low mechanical 
and thermal drift of the ring cavity geometry and is conceived to be operative in two different orientations of the laser plane, in order to detect rotations
around the vertical or the horizontal direction. 
Since June 2010 the system is active inside the Virgo interferometer central area with the aim of performing high sensitivity measurements of environmental  rotational noise.
So far, continuous unattended operation of the gyroscope has been longer than 30 days. 
The main characteristics of the laser, the active remote-controlled stabilization systems and the data acquisition techniques are presented. An off-line data processing, supported
by  a simple model of the sensor, is shown to improve the effective long term stability. A rotational sensitivity at the level 
of $10^{-8}~\rm{rad/\sqrt{Hz}}$ below $1~\rm{Hz}$, 
very close to the required specification for the improvement of the Virgo suspension control system, 
is demonstrated for the configuration where the laser plane is horizontal. 
\end{abstract}
\section{Introduction}
Laser gyroscopes (``gyrolasers'') are extremely sensitive absolute-rotation sensors based on the Sagnac effect. Two counterpropagating laser  beams  oscillate  on the same 
mode of an active ring laser cavity. When the cavity rotates with respect to an inertial reference frame, the optical frequencies of the two intracavity laser beams 
become different. 
The  angular velocity $\vec{\Omega}$ of the laser reference frame is related to the frequency difference $\Delta f$ between the two optical beams by the following relation:
\begin{equation}
\Delta f = \frac{4 A}{\lambda P}\vec{n} \cdot \vec{\Omega}, 
\label{ideal}
\end{equation}
where $A$ is the area enclosed by the optical path inside the cavity, $P$ the perimeter, $\lambda$ the optical wavelength, and $\vec{n}$ the area versor. 
To measure the frequency difference $\Delta f$, the two beams are superimposed on a photodetector and their beat note is detected. Gyrolasers find application in different fields like  navigation, 
geophysics and geodesy. 
The large frame ring laser gyroscope ``G'' \cite{schr09}, a monolithic square cavity 4~m in side, located by the Geodetic  Observatory of Wettzell (Germany),  routinely performs
rotation measurements with a resolution of the order of  few $\rm{prad/s}$. ``G'', together with other large gyrolasers around the world (for a detailed list of references 
see for example \texttt{http://www.ringlaser.org.nz}), provides 
essential information on rotational seismology \cite{rotational_seismology} and its long--term monitoring of the earth rotation rate makes the direct observation 
of geodetic effects like the solid earth tides \cite{schr03} and the diurnal polar motion \cite{schr04} possible. The present  sensitivity level is not too far 
from what is required for ground based General Relativity tests, 
which seem achievable in the near future by improved devices \cite{sted97,sted03,noilense}.

In this paper we present the experimental results concerning the use of a meter size gyrolaser as very sensitive tilt sensor.
Recently, this opportunity has  been considered by the community of the large  gravitational-waves interferometers like Virgo and Ligo with the 
aim of performing seismic monitoring and improving the control of the inertial suspensions. 
The required sensitivity for the next generation of the Virgo antenna is at the level of 
$10^{-8}~\rm{rad/\sqrt{Hz}}$ in the range $5-500~\rm{mHz}$ (see: The Virgo Collaboration, 2009 Advanced Virgo Baseline Design, Virgo note VIR-027A-09 (26 May 2009)). 

In the context of the experiment ``G-Pisa'' \cite{ncb}, a square-cavity gyrolaser,  $1.35~\rm{m}$ in side, has been assembled and has already demonstrated
a  sensitivity  at the level of few $\rm{(nrad/s)}/\sqrt{Hz}$ in the range  $10-100~\rm{mHz}$ \cite{CQG}. 
Since June 2010, ``G-Pisa'' worked inside the Virgo central area, few meters apart from the beam-splitter
and the two input mirrors of the 3 km Fabry-P\'erot cavities, with the main purpose of performing in situ measurements
 of the tilt motions, which are  responsible of the reduction of the antenna sensitivity during severe weather conditions.
Differently from  ``G''\cite{schr09}, the best performing ring-laser, which is based on a monolithic design and made in Zerodur (a glass with ultra-low expansion
coefficient at room temperature), our design is based on a largely more economical vacuum chamber in  stainless steel \cite{geosensor1}, rigidly fixed on a slab of granite.

\begin{figure}
\resizebox{1\textwidth}{!}{%
\includegraphics{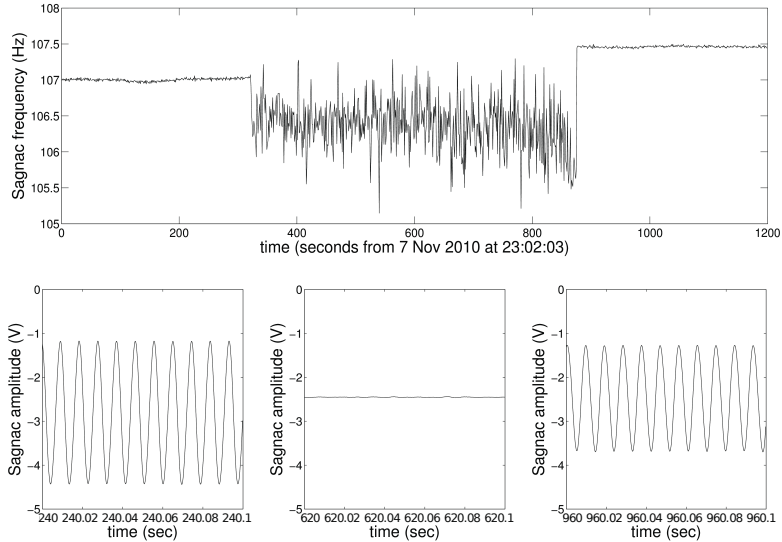}
}
\caption{Typical free-running behavior of the laser gyroscope. Upper trace: 
frequency tracking of the Sagnac beat signal. At $\rm{t=321~s}$  a mode jump brings the laser to ``split-mode'' regime. 
When the laser is split-mode the contrast  of the Sagnac interference falls to zero and the frequency detection algorithm returns noise. A second 
mode jump (at $\rm{t=877}$) recovers the ``standard'' operational regime.  The lower graphs show the Sagnac beat signal 
 before the mode jump (left), during the split-mode regime (middle)  and after the recovery of standard operation (right).}
\label{freerunning}       
\end{figure}

{Comparing the performances of G and  G-Pisa, we observed that, in the frequency range where both are shot-noise dominated, 
the sensitivity limit scales properly with the scale factor, the Q-factor and the laser power. However, in the very low  frequency 
band ($<10^{-1} \rm{Hz}$) G-Pisa is affected by some excess noise due to considerable differences in local rotation noise (tilts) and 
 environmental parameters stability (temperature and pressure) \cite{sted97}. Environmental temperature variations induce deformations of the laser cavity shape and therefore    
 change the calibration coefficient (scale factor) connecting the measured Sagnac frequency to the rotation rate.} In single-mode 
free running operations, a change in  the perimeter length produces a change in the laser wavelength. 
When the corresponding optical frequency variation  becomes comparable with the cavity free spectral range ($\sim55.5~\rm{MHz}$),  mode jumps may occur, bringing the laser to an unstable working regime. 
Moreover, mode jumps do not happen at the same time for the two counter-propagating beams; 
there is in general a time gap, that can be of the order of several minutes, 
where the two opposite beams work on two different longitudinal modes separated by one FSR  (split-mode operational regime). In this regime,
 the Sagnac signal shifts around the FSR frequency and the information about rotation is lost \footnote{The recovery of the rotational signal in split-mode regime  
can be obtained by heterodyning the beat signal with a stable oscillator at the FSR frequency. This solution is typically 
employed in ultra-large ring lasers \cite{ULTRA}.}.
A typical waveform of the Sagnac interference signal during a series of  split-mode transitions is sketched in Fig.\ref{freerunning}.

A control system has been implemented, in order to avoid mode jumps, by keeping  the perimeter of the ring constant
against the wavelength of a frequency-stabilized laser. This is obtained by moving 
two of the four mirrors along a diagonal by piezoelectric actuators. This stabilization system has a further advantage:  since the detuning of the
two lasers is absolutely fixed with respect to the atomic gain line, the frequency dependence of the non-linear atomic absorption  and dispersion effects can be considered constant for the two 
opposite intra-cavity beams. Amplitude stabilization of the laser emission power is as well necessary to ensure stable operation, and compensate for degradation of the gas mixture or for electronic 
fluctuations in the discharge excitation; in our device the power of one of the two modes is stabilized. 

In the following we will present a simple model  of the sensor (see for example: A. Velikoseltsev, Ph.D. Thesis, Tech. Univ. Muenchen, Germany, 2005) 
taking into account the main systematic effects in the rotation measurement. We will describe the experimental apparatus which is now operating inside the Virgo
central area, giving the details of the stabilization loops, and of the remote control system, 
which is necessary since most of the time the access to the Virgo central area is forbidden. Finally, the experimental data treatment and analysis will be  discussed, 
focusing on the sensitivity to angular movements in the frequency region below 100~mHz down to 1~mHz.

\section{Sensor Model}
\label{modello}
The relation expressed in Eq.(\ref{ideal}), which connects the beat note with the angular velocity and the geometrical parameters of the ring, 
 is only a first order approximation.

A more complete model of the ring laser  gives the following expression for the measured beat frequency $\Delta f$:

\begin{equation}
\Delta f = K_R (1 + K_A )\vec{n}\cdot \vec{\Omega} + f_0 + f_{bs}.
\end{equation}

The first term is the scale factor, the second, $f_0$, is the null-shift error
and the last, $f_{bs}$, is given by backscattering effects. The coefficients of the
scale factor are defined as follows:
$K_{R} = 4A/(\lambda P)$ is the geometrical scale factor that links the Sagnac effect to the geometry of the laser resonance cavity. 
It must be corrected by the coefficient  $K_{A}$ which takes into account the changes in the optical path length due to the dispersion properties of the plasma 
in the discharge region.

{The non-reciprocity of the optical cavity is the  source of the null-shift error $f_{0}$. 
Non-linear effects in the plasma arising from parity non-conserving interactions ( 
such as the Faraday rotation combined to a slight elliptical polarization) can be kept
 under control by accurately shielding the discharge from stray magnetic fields.
In addition, environmental factors like  temperature and pressure fluctuations in the laboratory, as well as outgassing, 
produce different losses for the two counter-propagating beams.} The null-shift  contribution, under frequency stabilized operations, 
turns out to be proportional to the single beams intensity difference $\Delta I=I_{+}-I_{-}$:
\begin{equation}
f_{0}=\alpha_{0} \Delta I.
\end{equation}
 Since  beam-power measurements are available for each beam, the intensity difference can be
taken directly from the experimental data and the evaluation of the null shift contribution can be obtained by fitting $\alpha_0$ in an off-line analysis.

The last term is produced by the backscattering of the radiation,
mainly on the mirrors, that couples the circulating beams one to the other.
{Since the backscattering cross-section is inversely proportional to the perimeter of the ring cavity  
 this is the dominant error contribution in the small ring laser gyros.}
Since the amplitude and the phase of the backscattered light affects also the laser intensity, 
an estimation of $f_{bs}$ can be obtained by implementing a diagnostic on the two single beam outputs of a single cavity mirror. In particular, we can write
\cite{brevetto}:
\begin{equation}
\label{discriminant}
f_{bs}\propto\mathcal{D}=\langle(D_{bs}-\langle{D}_{bs}\rangle_{T})^2\rangle_{T}
\end{equation}
where $D_{bs}=\frac{P}{c}(\frac{\dot{I}_+}{I_+} + \frac{\dot{I}_-}{I_-})$, and where the symbol $\langle \rangle_{T}$ indicates the average operation calculated over a 
time interval ${T}$, containing many cycles of the Sagnac beat oscillation signal. 
The proportionality constant between $\mathcal{D}$ and $f_{bs}$ is then fitted to the experimental data. Details are reported in the appendix.

\section{Experimental apparatus}
\begin{figure}[h!!!]
\resizebox{1\textwidth}{!}{%
\includegraphics{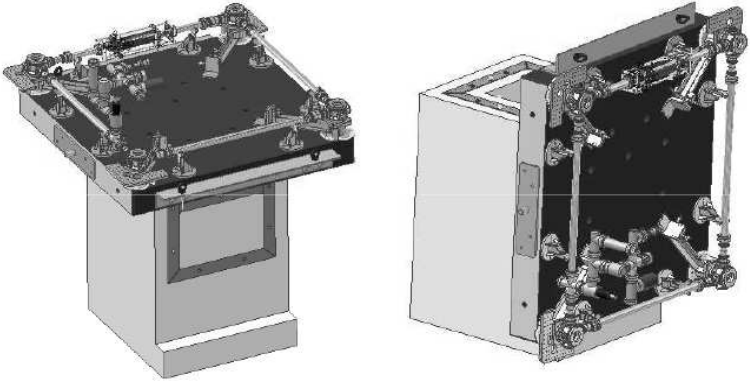}
}
\caption{Mechanical design of gyroscope in the two possible orientations of the laser plane.}
\label{CAD_bosi}       
\end{figure}

The mechanical design of the laser gyroscope is sketched in Fig. \ref{CAD_bosi}. 
A $180~\rm{mm}$ thick and $1.50~\rm{m}$ in side square granite slab supports the 
whole mechanical ring and defines the laser cavity reference frame. 

The Virgo experiment is mainly interested in the ground rotation around the horizontal axes; 
for this reason a steel armed reinforced concrete support has been designed and realized, which is able to support 
the granite table both horizontally and vertically, in order to measure the rotations around the vertical, or around the horizontal direction. 
A steel flange is  embedded at the center both of the upper side and of the lateral side of the concrete monument,  
in order to firmly hold the granite table. The weight of the concrete support is $2.2 ~\rm{ton}$, while the granite table is around $1~\rm{ton}$. 
The weight of the whole structure has to guarantee a good contact with the underneath floor. 
In order to improve  this contact as much as possible, a liquid,  fast-setting, concrete has used to fill cracks and gaps between the floor and the monument basis.
The optical cavity design is based on the GEOsensor \cite{geosensor} project. As stated before, it is a 
 square optical cavity, 5.40~m in perimeter and 1.82~m$^2$ in area, enclosed in a vacuum chamber entirely filled with the active medium gas.
The  vacuum chamber has a stainless steel modular structure: 4 boxes, 
located at the corners of the square and containing the mirror holders inside, are connected by pipes through flexible bellows, 
in order to form a ring vacuum chamber with a total volume of about $5\cdot 10^{-3}$~m$^3$. 
The mirrors are rigidly fixed to the boxes. The mirrors alignment can be achieved thanks to  
a micro-metric lever system that allows to regulate the two tilt degrees of freedom of each box. A fine movement of two opposite placed boxes 
along the diagonal of the square is also possible. This is provided by two piezoelectric transducers that allow the servo control of the laser cavity perimeter length. 
No window restricts the active region and the vacuum chamber is entirely filled with a mixture of He and a $50\%$ isotopic mixture of $^{20}$Ne and $^{22}$Ne. 
The total pressure of the gas mixture is set to $560~\rm{Pa}$ with a partial pressure of Neon of $20~\rm{Pa}$. 
The active region is a plasma which is generated by  a RF capacitive discharge coupled to the gas through a pyrex capillary inserted at the middle of one side 
of the ring. 
The ring cavity mirrors have  a radius of curvature of 4~m.
By abruptly switching off the discharge excitation, 
we measured a shut-down decay time of the laser radiation of  $0.66~\rm{ms}$, giving  an effective optical cavity quality factor 
$Q=2\times10^{12}$, consistent with a mirror reflectivity of $99.9992\%$. The laser is working very close to threshold, 
in order to select only a single longitudinal mode ($TEM_{00n}$). The typical  power of a single output beam is typically of the order of $10~\rm{nW}$.

\section{Perimeter digital control}
\begin{figure}
\resizebox{1\textwidth}{!}{%
\includegraphics{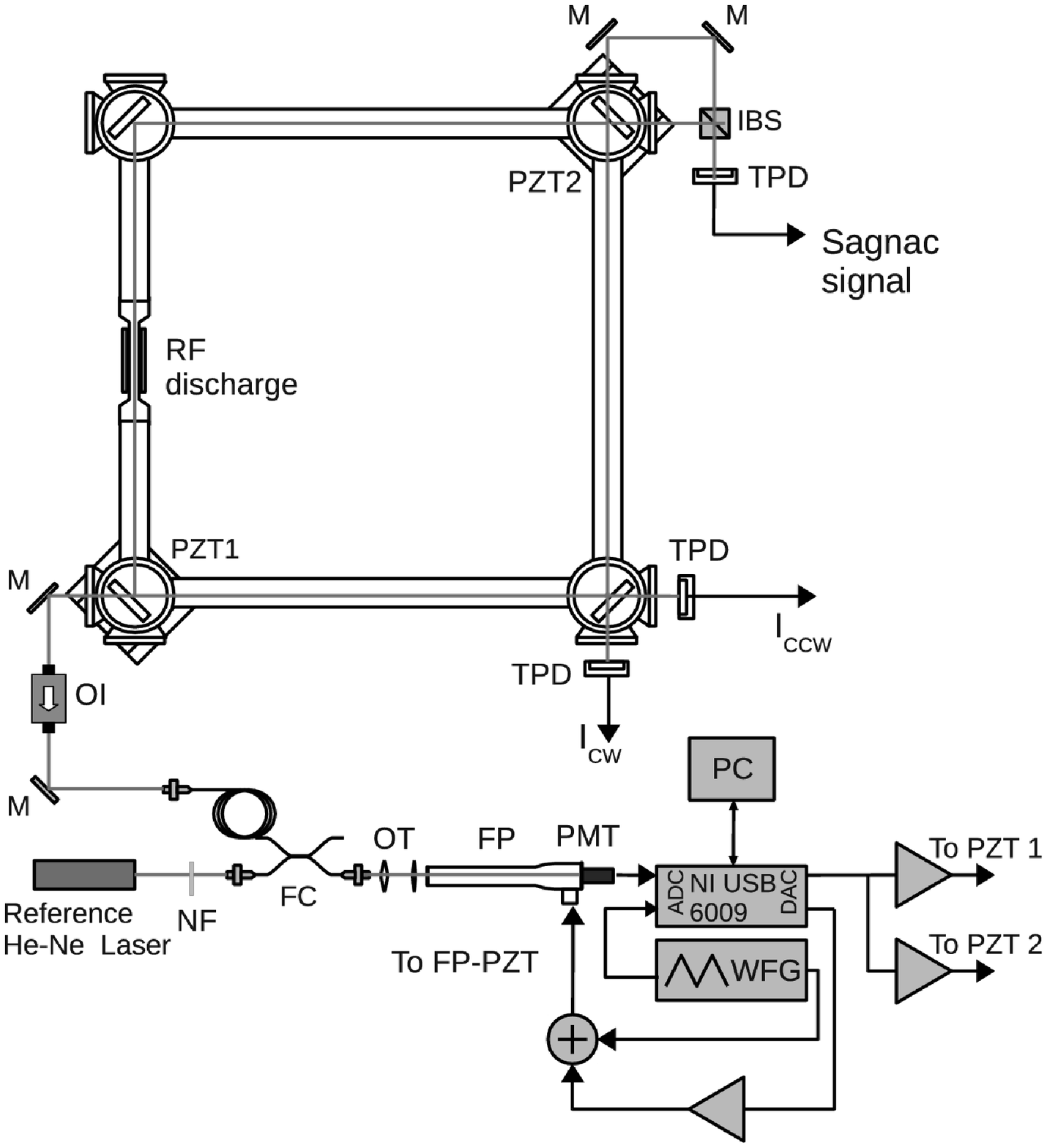}}
\caption{Experimental set-up of the perimeter-controlled gyrolaser. The optical elements for the detection of the rotation signal (Sagnac signal) and  of the clockwise 
and counterclockwise intensities ($I_{+}$ and $I_{-}$) is also shown. FP: Fabry-P\'erot analyzer, M: mirror, IBS: intensity beam splitter, TPD: transimpedance photodiode, 
PMT: photomultiplier, OI=optical isolator, FC: fiber coupler, NF: neutral filter, OT: optical telescope,  PZT: piezoelectric transducer, WFG: waveform generator.}
\label{Cavity2PZT}
\end{figure}

The long term perimeter control is obtained by comparing the gyrolaser optical frequency with a He-Ne reference laser which is frequency-stabilized to the Doppler 
broadened profile of the laser transition. The correction is applied to the ring cavity by acting on the  piezoelectric devices moving the mirrors boxes. 
The long term stability of the reference laser  is given of the order of $1-2~\rm{MHz}$ over one year \cite{bob}. The frequency separation between the two lasers
 is measured by means of a 
Fabry-P\'erot spectrum analyzer \cite{biancalana} and the ring laser perimeter length is corrected in 
order to keep this difference equal to a constant value around $60~\rm{MHz}$, which corresponds to the effective maximum of the  gain curve as determined 
by the superposition of the Doppler broadened gain curves of $^{20}\rm{Ne}$ and $^{22}\rm{Ne}$.
The control scheme is sketched in Fig.\ref{Cavity2PZT}. Both the radiation emitted from the gyrolaser and the reference laser are injected into an 
optical fiber and superimposed in a $2\times2$~fiber--combiner. The output of the fiber coupler is mode-matched to a scanning Fabry-P\'erot analyzer (FP)
 with a free spectral range of $300~\rm{MHz}$, and a finesse of about 100; the transmitted intensity is detected by a photomultiplier. The FP cavity length
 is continuously scanned  by driving the piezoelectric transducer with a triangular waveform at the frequency of $2~\rm{Hz}$. 
Initially, the central value of the frequency scan, set by the offset of the triangular waveform, is tuned at half way between the reference laser and the gyrolaser 
optical frequency. 
After each scan the FP optical spectrum, containing the two resonance peaks, is processed by a PC. 
The positions of the two peak centers, the one of the gyrolaser and the one of the reference laser, are estimated via a parabolic fit of the data around the two 
transmission maxima.
Once the resonances positions are estimated, a double digital PID feedback loop, acting on the gyrolaser cavity length and on the offset 
voltage of the FP PZT, is then implemented by means of two independent DAC channels\footnote{Both PIDs have a
predominant integral behavior with an integral gain of about 1. The proportional and derivative gains have much lower values
and are properly set in order to improve the loop stability.}.
 The digital control system can be operated from remote, via a VNC connection, during the periods when the Virgo 
central area is not accessible.

The perimeter control implements a correction acting on the position of two opposite mirrors of the cavity.
When the perimeter is locked to a fixed value, the geometry of the ring is distorted and the area accordingly changes. 
However, considering that we are very close to the perfect square condition (construction mechanical tolerances of about 1~\rm{mm}), 
the relative change in the diagonal does not produce a sensitive relative change of the Sagnac frequency. 
Since the temperature dependence of the perimeter length was estimated as $\sim30~\rm{\mu m/K}$, the error in the rotation rate 
estimate due to the geometrical deformation of the cavity is at the level of $\rm{3~(prad/s)/K}$. 
The offset voltage of the FP ramp is actively controlled in order to keep the position of the reference laser resonance peak, with respect to
 the starting value of the ramp, constant. This allows to compensate the thermal drift of 
the analyzer cavity length. This choice allows also to perform narrow sweeps around the two resonances in order to improve the frequency 
resolution and the correction-loop velocity. It is worth noting that the length of the FP analyzer does not influence the accuracy of the frequency spacing determination, 
because it determines a common mode shift of the resonance peaks of the two laser which have only a slight frequency difference. 
Using a single PZT to perform both frequency scans of the FP and correct its central value is a very practical method. This method
 presents however some intrinsic limitations connected to the non-linearity of the PZT response. A refined version of the method, improving the long term stability
of the gyrolaser frequency would involve the use a FP cavity having two separate piezos separately dedicated to scan and to tune the DC level of the FP length, 
or to use a thermostatic technique \cite{Orozco}.

\subsection{Backscattering phase and double PZT control scheme}

A first implementation of the perimeter control has been described in \cite{ncb}. It was applied to the same ring structure, but with a slightly different side dimension (1.40 m). 
In that case, the correction was made by acting on a PZT at a single mirror.

As already observed in \cite{Rodloff} and \cite{ulli_precision}, this operation solves the problem of the 
discontinuous operation regime characterized by “split-mode” transitions,
but introduces a slow modulation in the value of the Sagnac frequency that
results to be proportional to the temperature drift rate of the gyrolaser
structure. This can be explained by considering that  moving  a single mirror along 
the cavity diagonal changes the length of the two adjacent sides, while the two opposite side are, at first order, unchanged. 
Thus, temperature drifts determine 
an asymmetric deformation of the cavity shape in closed loop conditions. As we will discuss further, this fact strongly affects 
the backscattering induced Sagnac frequency pulling \cite{sted97,harm_anal}. 

\begin{figure}
\resizebox{1\textwidth}{!}{%
\includegraphics{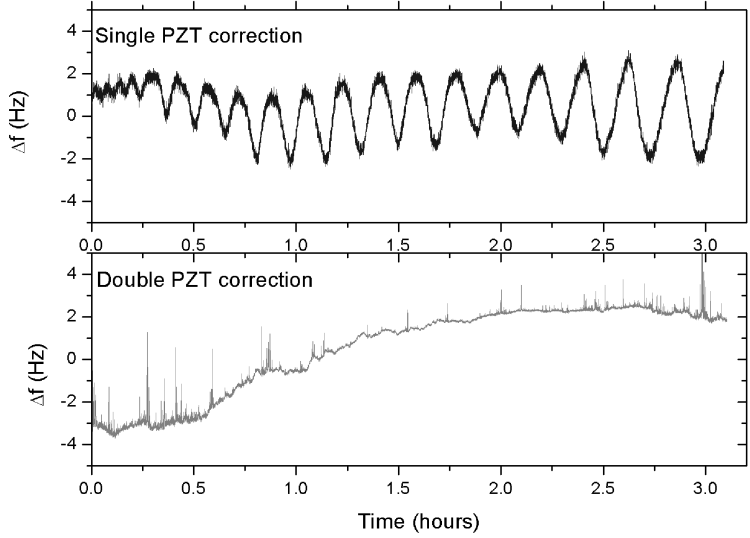}}
\caption{Two sets of three hours trends of the Sagnac frequency when the perimeter control system is engaged. The upper trace refers to the case where 
one single movable mirror is used to compensate the perimeter variations due to thermal and mechanical drifts.  The lower trace refers to the double PZT scheme. 
In both cases the room temperature was drifting with a rate  of about $0.2~\rm{K/h}$. The spikes observed in the lower trace are an electronic artifact
 produced by the digital frequency recognition algorithm, when the  signal-to-noise-ratio of the Sagnac beat signal is too low.} 
\label{cnfr_singPZT_double_PZT}
\end{figure}

The solution we have employed is to control the cavity perimeter by using two PZTs, which change the length of one diagonal of the square,
without moving its geometrical center. 
{This technique turns out to be efficient to increase the stability of the rotational signal against  temperature fluctuations.  
The double PZT scheme, in the case of a perfectly aligned cavity, would completely 
cancel the temperature coefficient of the total backscattering phase, and then of the frequency pulling, against isotropic deformations of the cavity.
 We characterized the performances of such a scheme on a former gyrolaser  set-up having a Q-factor about 30 times lower than in the present setup. 
The amount of backscattering-induced frequency pulling was in that case at the level of a few Hz. A residual dependence on temperature was experimentally 
observed also in this configuration mainly due to   cavity misalignments,  
anisotropic deformations of the laser frame (induced, for example, by temperature gradients) and the residual asymmetry of 
the two piezo elements. A comparison between the behavior of the 
Sagnac frequency when the perimeter is stabilized actuating one single 
mirror (upper graph) and when the double PZT control system is engaged (lower graph) is shown in Fig.\ref{cnfr_singPZT_double_PZT}. 
In the first case the backscattering phase changes at the  rate of about $20~\rm{cycles/K}$. 
In  the second case  the backscattering phase change rate is reduced by a factor of about 40.}

\section{Amplitude stabilization}
The intensities of the two counterpropagating  beams are influenced by several factors. An amplitude modulation at the Sagnac frequency is present, in general, 
superimposed on the DC level. Such modulation is produced by the interference between the photons of one beam with the backscattered photons of the other, 
frequency shifted, one. The long term intensity stability is, on the other side, affected by: the  optical misalignments in the light path, due to thermal effects 
and mechanical settlings; the variation of the RF power discharge, caused by aging and self-heating of the electronic components;
the variations in the composition of the gas inside the cavity (hydrogen contamination), changing  the properties of the plasma and the optical gain.

\begin{figure}[h]
\resizebox{1\textwidth}{!}{%
\includegraphics{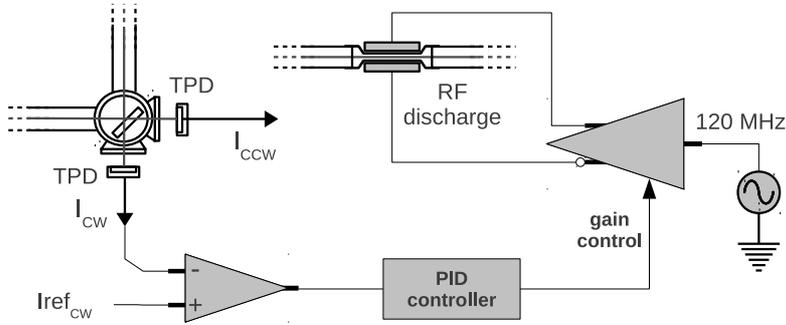}}
\caption{Block diagram of the amplitude stabilization system.}
\label{stabilizzazione}
\end{figure}

As a first consideration, there are sharp upper and lower limits for the beam intensities that should not be exceeded: 
the upper limit corresponds to the transition to the multimode regime, while the lower limit is the laser threshold. Minor variations of the beam intensities (other than those related to the Sagnac effect) also should be avoided: 
they induce undesired optical gain modulation and non-linear optical dispersion effects.
In order to avoid such undesired intensity fluctuations we have implemented a closed-loop stabilization system based upon a PID analog controller,
 the scheme of which is shown in Fig. \ref{stabilizzazione}.
The inputs of the system are the intensity of one beam (clockwise or counter-clockwise) as revealed by 
the photodiode trans-impedance amplifier, and an externally imposed reference intensity value. The output voltage of the PID controller 
sets the power of the RF plasma excitation through a variable gain RF amplifier, thus obtaining the desired intensity of light emission 
and closing the loop. 
\footnote{The settling time of the closed-loop system can be manually adjusted in the range 0.2-2~s,
therefore the active control does not operate at the Sagnac frequency.}

\begin{figure}[h]
\resizebox{1\textwidth}{!}{%
\includegraphics{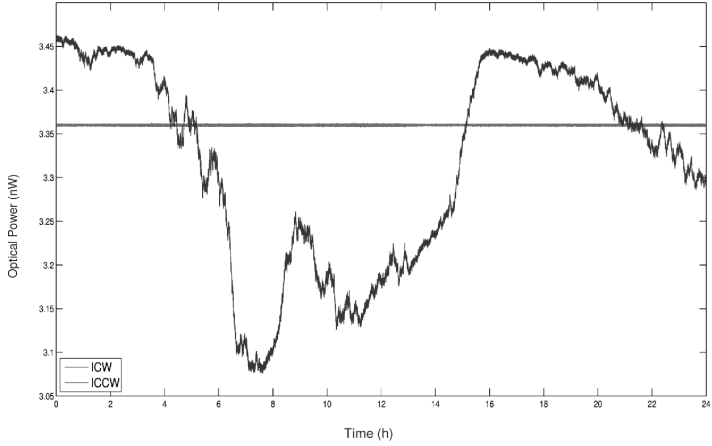}}
\caption{Output power of the CCW and CW beam, when the CW intensity stabilization loop is engaged. 
A different level for the two single beam output intensities is justified by the non-reciprocal phenomena happening in the active medium 
and by the non reciprocal optical losses. When operating under frequency stabilized conditions the dishomogeneity of the mirrors surfaces plays the dominant role.}
\label{Fabio1}
\end{figure}

Figure \ref{Fabio1} shows a plot of the CW and CCW intensities during a period of 24 hours, the beam employed in the stabilization loop is the CW one.
This laser amplitude stabilization reduces the long-term fluctuations, 
and increases the
duty cycle of the apparatus, avoiding multi-mode and extinction of the laser beam. However, this kind of technique is not ideal, as 
demonstrated by the large fluctuation observed in the counterpropagating beam.
Some different approaches could be tested, by combining  the two intensity error signals in different ways and on different time-scales.
\section{Data acquisition and analysis}
The data from the gyrolaser are acquired and stored continuously by the Virgo data acquisition system. The Sagnac beat signal and the two single beam intensities
are acquired at the rate of $5~\rm{kSample/s}$ so that it is possible to reconstruct the Sagnac  phase-noise nominally up to $2.5~\rm{kHz}$.
In addition, a local PC provides the evaluation of the instantaneous Sagnac frequency as well as of   $\Delta I$ and $\mathcal{D}$ (see section \ref{modello}) 
at the rate of $1~\rm{Hz}$ in order to provide
real time estimations  of rotation rate and of the relevant parameters for the off-line correction of the systematic effects introduced by the laser dynamics.
Since we are interested in the long term monitoring of local rotations, at the level of $\rm{(nrad)/s}$, which are superimposed on the Earth rotation-rate bias, 
it is fundamental to use a clock with an excellent long term stability for  the data acquisition timing process. The Virgo GPS-based central timing
system \cite{Virgo_Timing}, developed to synchronize the Virgo interferometer’s controls and readouts, is employed. 

\begin{figure}[h]
\resizebox{1\textwidth}{!}{%
\includegraphics{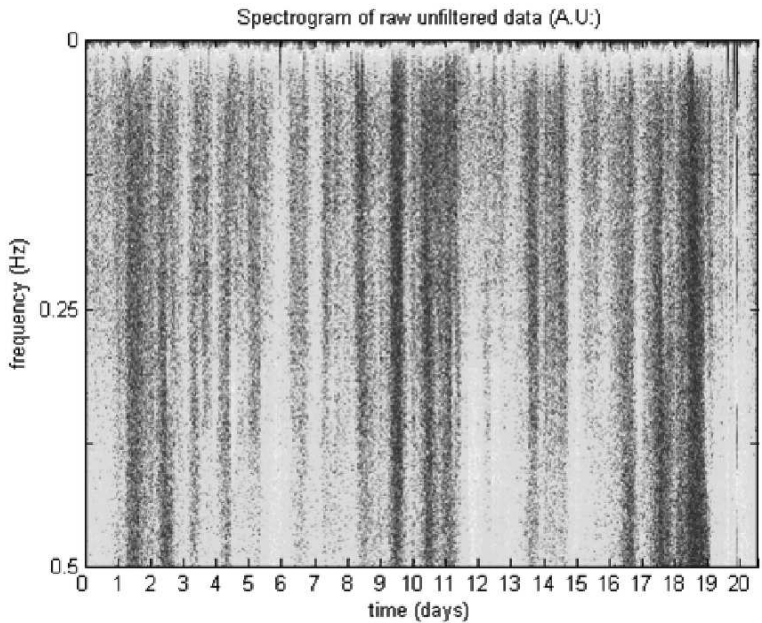}}
\caption{Spectrogram of the power spectrum of the rotational noise, the day-night cycle as well as the  weekly are observable.}
\label{Immagine3}
\end{figure}

``G-Pisa'' has been running unattended for several months with a vertical axis orientation. Fig.\ref{Immagine3}  shows the spectrogram 
of the Sagnac frequency for a time period of 20 days, starting from the beginning of August 2010. The  noise level  is modulated by the
night/day cycle, quieter at night and noisier in day time. A longer modulation of the noise level, at  weekly time scale, is also observable.
The average Sagnac frequency is estimated from the same dataset as  $107.32~\rm{Hz}$.
After correcting the data from the systematic effects, given by backscattering and intensity difference, the average value becomes $107.37~\rm{Hz}$.
This value has to be  compared with the expected value of $107.47~\rm{Hz}$ calculated for a horizontally  positioned gyro, 
located at latitude of $43^{\circ} 37^{'} 53^{''}$. The measured and the expected values are consistent admitting 
 an angular error smaller than $1~\rm{mrad}$ in the alignment between the laser area vector and the local  vertical direction.
  
\begin{figure}[h]
\resizebox{1\textwidth}{!}{%
\includegraphics{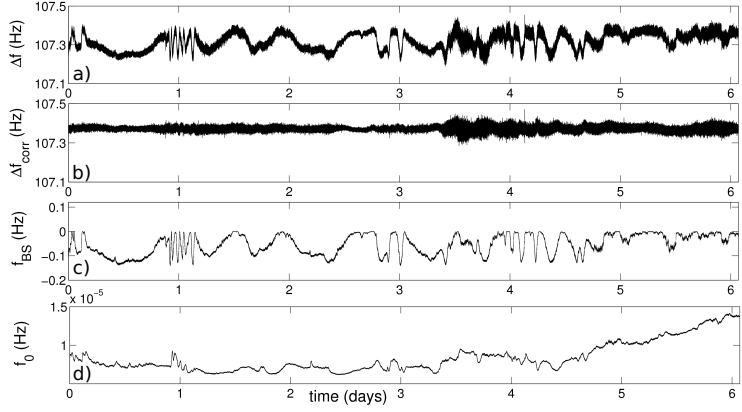}}
\caption{a: 6 days monitoring of the earth-rate induced Sagnac frequency from 10:00~am of the 22th of November 2010; 
b: residual fluctuations after the correction for the backscattering and null-shift errors;  c: best fit result for the backscattering induced frequency pulling
d: best fit result for the null shift contribution.}
\label{threeFig}
\end{figure}

In the first run the laser power  control was unstable, and   
affected the response of the instrument around and below $100~\rm{mHz}$. This problem has been fixed in the middle of November, and the following analysis refers to the 
period between the 17th of November and the 17th of December, when the system was shut-down in order to change the orientation of the ring and to be sensitive to the tilt
along the horizontal axis.
To improve the low frequency sensitivity we have done the attempt to subtract the noise coming from $\Delta I$ and $D_{bs}$. 
Fig. \ref{threeFig} shows the time series relative to the Sagnac frequency, the backscattering contribution and the relative power difference of the two 
counterpropagating modes. 

\section{Sensitivity estimation}
\begin{figure}
\resizebox{1\textwidth}{!}{%
\includegraphics{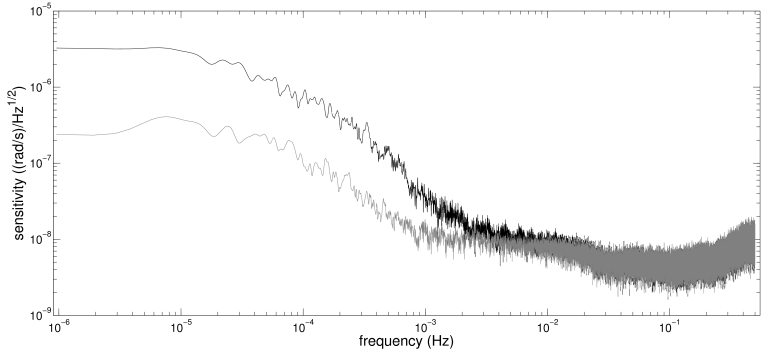}}
\caption{Rotation rate sensitivity with and without the subtraction of the systematic contributions from null-shift and backscattering induced 
frequency pulling.}
\label{BackSub}
\end{figure}

The rotation-rate  sensitivity limit of the actively stabilized gyroscope, with and  without the subtraction of the backscattering 
and the power difference effects, is shown in Fig. \ref{BackSub}. 


\begin{figure}
\resizebox{1\textwidth}{!}{%
\includegraphics{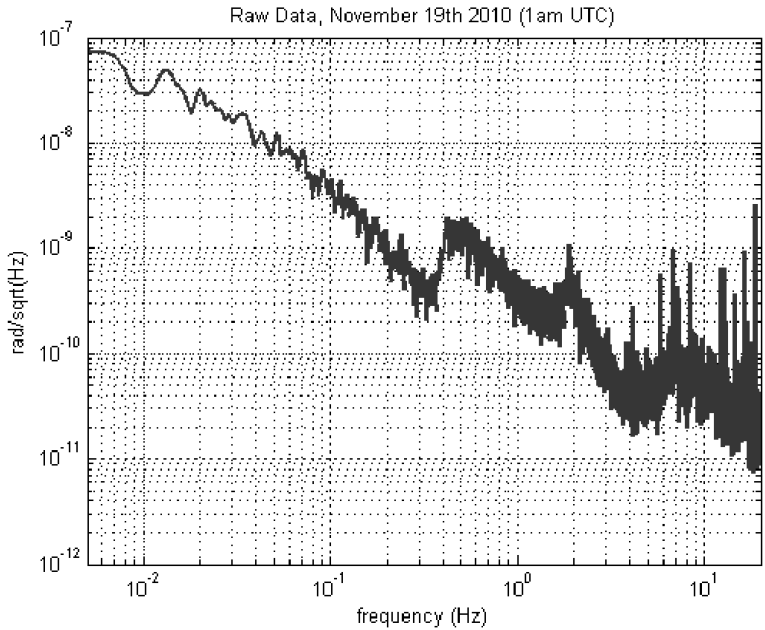}}
\caption{Tilt sensitivity in a very quiet period, the requirements of AdVirgo are $10^{-8}~\rm{rad/\sqrt{Hz}}$ in the range $5-500~\rm{mHz}$.}
\label{Immagine1}
\end{figure}

The off-line correction of the systematic effects becomes effective below 
the frequency of $10~\rm{mHz}$, and provides a gain in sensitivity of about one order of magnitude around $100~\rm{\mu Hz}$. 
A tilt sensitivity curve, expressed as $\rm{rad/\sqrt{Hz}}$ referred to a time period of very low environmental noise level,
is presented in  Fig. \ref{Immagine1}. 

\section{Conclusions}

In this paper we characterized a meter-size gyrolaser dedicated to the monitoring of the rotational noise of the Virgo interferometer. The system is
completely active-controlled, in frequency and power. The control system makes it possible to avoid mode jumps and multimode regimes
for months and consequently to operate the laser without accessing the Virgo area over the interferometer data taking periods.  

A new configuration of the  perimeter control, moving two opposite cavity mirrors symmetrically 
along the cavity diagonal, was developed.  With this new scheme we demonstrated a strong reduction 
of the backscatter contribution with respect to the 
 configuration where a single movable mirror is used. This leads to an effective increase of the  
stability of the rotational signal against the environmental temperature fluctuations. It is worth noting that the use of two PZTs moving 
different mirrors provides the opportunity to act on two degrees of freedom: the total backscattering phase and the cavity perimeter.

A data acquisition system, integrated into the Virgo scheme, has been developed. A  continuous sampling of the Sagnac interference signal and of the 
two single-beam intensities at the sampling frequency of $5~\rm{kHz}$, allows for a very accurate estimation of the systematic effects.
Off-line refinement of the rotational signal, performed by subtracting the systematic frequency shifts given by backscattering ($f_{bs}$) and
non reciprocities ($f_{0}$) is shown to produce an improvement in the low frequency sensitivity of the instrument, of about 
one order of magnitude below 1~mHz.
 
The final measured sensitivity limit to the tilts around the vertical axis, expressed as $\rm{rad/\sqrt{Hz}}$, around $10~\rm{mHz}$ 
is at the level of  $10^{-7}~\rm{rad/\sqrt{Hz}}$,  which is close to what is required for AdVirgo.

\section{Appendix}
We report here a derivation of the quantity used to estimate the backscattering-induced frequency pulling of the Sagnac interference signal.
The equations for the intensities $I_{+}$ and $I_{-}$ and the phase difference $\psi$ of the clockwise (+) and the counterclockwise (-) laser beams are
 written as \cite{sted97}:
\begin{eqnarray*}
\frac{1}{f_{FSR}} \frac{\dot{I}_{\pm}}{I_{\pm}}&=&(\alpha_{\pm}-\beta_{\pm} I_{\pm} - \theta_{\pm} I_{\mp})+2 r_{\pm}\sqrt{\frac{I_{\mp}}{I_{\pm}}}\cos{(\psi \pm \zeta)},\\\nonumber
\dot{\psi}&=& \omega_S+(\sigma_{+}-\sigma_{-})+(I_{+}\tau_{-}-I_{-}\tau_{+})\\\nonumber
&&-f_{FSR}[r_{-}\sqrt{\frac{I_{-}}{I_{+}}}\sin{(\psi+\zeta)}+r_{+}\sqrt{\frac{I_{+}}{I_{-}}}\sin{(\psi-\zeta)}],
\end{eqnarray*}
where: $f_{FSR}$ is the free-spectral-range of the laser cavity, $\omega_S$ is the difference between the resonant frequencies for the two counterpropagating beams,  
$\alpha_{+,-}$, $\beta_{+,-}$, $\theta_{+,-}$, $\sigma_{+,-}$, $\tau_{+,-}$  are the coefficients that describe the active medium dynamics. 
They are functions of the Neon isotopic ratio, the optical  frequencies of each of the traveling waves and the total gas pressure.
The coupling terms are due to the cross saturation of the laser transition  and to the backscattering. $\theta_{+,-}$ is the mutual gain coupling, 
$\tau_{+,-}$ is  the mutual dispersion coupling and $r_{\pm}$ is the backscattering coupling of each of the beams into the other with phase angles $\epsilon_{+,-}$. 
$\zeta=\frac{\epsilon_++\epsilon_-}{2}$ is the net backscattering phase.
In the case where $r_{+}\simeq r_{-}=r$, $I_+\simeq I_-$,  $\sigma_{+}\simeq\sigma_{-}$, $\tau_{+}\simeq\tau_{-}$, the equation for the Sagnac phase becomes:
\begin{eqnarray}
&&\dot{\psi}=\omega_S -2 r f_{FSR} \cos{(\zeta)}\sin{(\psi)}.
\end{eqnarray}
The above equation, for $\omega_S>2 r f_{FSR}  \cos{(\zeta)}$ has the following analytical solution: 
\begin{eqnarray}
\psi(t)=2 \arctan\big\{{\frac{\Omega_L}{\omega_S}+\frac{\sqrt{\omega_{S}^2-\Omega_{L}^2}}{\omega_S}\tan\big[{\frac{1}{2}(t-t_0)\sqrt{\omega_{S}^2-\Omega_{L}^2}}}\big]\big\}
\end{eqnarray}
where:
\begin{equation}
\label{pulled_freq}
\omega_p=\sqrt{\omega_{S}^2-\Omega_{L}^2}
\end{equation}
is the pulled Sagnac frequency by backscattering effects, being $\Omega_L=2 r f_{FSR} \cos{(\zeta)}$  the lock-in frequency. 
Under the same approximations it is useful to consider the observable $D_{bs}$:
\begin{eqnarray}
D_{bs}&=&(\frac{\dot{I}_+}{I_+} + \frac{\dot{I}_-}{I_-})\\\nonumber 
&=&K - 2 r  f_{FSR}[\cos{(\psi+\zeta)}+ \cos{(\psi-\zeta)}]\\\nonumber
&=&K-2 \Omega_L \cos{\psi}
\end{eqnarray}
where $K\simeq2[\alpha+(\beta+\theta)I] f_{FSR}$ is a nearly constant contribution coming from the atomic absorption/dispersion effects.
Remembering that $\psi$ is varying at the Sagnac frequency, it is possible to evaluate the contribution from the backscattering induced pulling by estimating the
amplitude of the modulation at the Sagnac frequency on $D_{bs}$ \cite{brevetto}. We consider then the following estimator:
\begin{equation}
\mathcal{D}=\langle(D_{bs}-\langle{D}_{bs}\rangle_{T})^2\rangle_{T}\propto \Omega_L^2,
\end{equation}
where the symbol $\langle\,\,\rangle_{T}$ indicates the average operation calculated over a 
time interval ${T}$.
The above quantity turns out to be proportional to the square of the lock-in threshold $\Omega_L$. Thus the error $f_{bs}$ in the Sagnac frequency estimation due to
the coupling between the two counterpropagating beams is obtained, to the first order in $\Omega_L/\omega_S$, as:
\begin{equation}
f_{bs}\simeq \alpha_{bs} \mathcal{D}. 
\end{equation}
The quantity $\alpha_{bs}$ is fitted once for each given measurement run of several days. 

\section{Acknowledgements}
We thank U.~Schreiber and A.~Velikoseltsev for the  very useful discussions. The installation inside the Virgo central building has  been possible 
only thanks to the strong support from the EGO and Virgo team. We would like to acknowledge M.~Bazzi, A.~Bozzi, S.~Braccini, E.~Calloni,   
F.~Carbognani, V.~Dattilo, G.~Di~Biase, F.~Frasconi, A.~Pasqualetti. 
We acknowledge G.~Balestri, G.~Petragnani and A.~Soldani from the INFN, section of Pisa.
We acknowledge M.~Francesconi, F.~Francesconi,  S.~Gennai, M.~Marinari, and R.~Passaquieti, from the Physics Department of Pisa.


%
%
%
%

\end{document}